\documentclass[10pt,conference,twocolumn,letterpaper]{IEEEtran}

\usepackage[english]{babel}

\usepackage{amsmath}
\usepackage{graphicx}
\usepackage[colorlinks=true, allcolors=blue]{hyperref}

\usepackage{multirow}
\usepackage{subcaption}
\usepackage{amsmath}
\usepackage{amsfonts}
\usepackage{amssymb}
\usepackage{algorithm}
\usepackage{algpseudocode}
\usepackage[acronym]{glossaries}
\usepackage[numbers]{natbib}

\newcommand{\argmin}{\arg\!\min}

\title{A Transformer Inspired AI-based MIMO receiver}
\author{\IEEEauthorblockN{András Rácz, Tamás Borsos, András Veres}
\IEEEauthorblockA{\textit{Ericsson Research} \\
Budapest, Hungary \\
\{andras.racz, tamas.borsos, andras.veres\}@ericsson.com}
\and
\IEEEauthorblockN{Benedek Csala}
\IEEEauthorblockA{\textit{Department of Broadband Infocommunications}\\ \textit{and Electromagnetic Theory}\\
\textit{Budapest University of 
Technology and Economics}\\
Budapest, Hungary \\
csala.benedekmihaly@edu.bme.hu}
}

\begin{document}
\maketitle

\begin{abstract}
We present \emph{AttDet}, a Transformer‐inspired MIMO (Multiple Input Multiple Output) detection method that treats each transmit layer as a token and learns inter‐stream interference via a lightweight self‐attention mechanism. Queries and keys are derived directly from the estimated channel matrix, so attention scores quantify channel correlation. Values are initialized by matched-filter outputs and iteratively refined. The AttDet design combines model-based 
interpretability with data-driven flexibility.  We demonstrate through link‐level simulations under realistic 5G channel models and high‐order, mixed QAM modulation and coding schemes, that AttDet can approach near‐optimal BER/BLER (Bit Error Rate/Block Error Rate) performance while maintaining predictable, polynomial complexity.
\end{abstract}

\section{Introduction}

Modern wireless communication systems demand ever‐increasing data rates and spectral efficiency, driving the adoption of multi‐antenna (MIMO) technologies in standards such as 5G and beyond.  By exploiting spatial multiplexing, multiple users or streams can be served simultaneously over the same time–frequency resource, dramatically boosting throughput.  
The difficulty of MIMO arises mainly in the detection algorithm at the receiver, which must disentangle the multiple spatial streams transmitted over the same time–frequency resource while contending with noise, interference, and channel correlations. The Maximum Likelihood (ML), ideal detector offers optimal error performance but its complexity grows exponentially with the number of layers, ruling out practical real-time implementations at scale. 

Linear detectors such as Zero Forcing (ZF) and Minimum Mean Square Error (MMSE) offer polynomial complexity and are widely used in practice but suffer performance loss in highly correlated channels, leaving a persistent gap to optimal. More advanced tree‐search detectors (e.g., K‐best) can approach ML performance but incur high and variable complexity, which limits practical applicability.


In response to these challenges, data‐driven approaches have emerged for MIMO detection.  \emph{Hybrid} methods such as DetNet \cite{song2019LearnToDetect} and MMNet \cite{shirkoohi2019adaptMassiveMIMO} “unroll” classical iterative solvers into neural networks, embedding domain knowledge to retain interpretability and reduce parameter count.  In contrast, \emph{end‐to‐end} or “black‐box” solutions—including DeepRx for MIMO \cite{korpi2020deeprxmimoconvolutionalmimo} and the Neural Receiver for 5G NR MU‐MIMO \cite{cammerer2023neuralreceiver5gnr}—leverage deep convolutional and graph neural networks to jointly learn channel estimation, equalization, and demapping. While these architectures can surpass traditional methods under various channel conditions, they suffer higher complexity and lack formal explainability.

Meanwhile, the Transformer architecture—originally developed for natural language processing—has demonstrated unprecedented ability to capture sequence-to-sequence learning via its self‐attention mechanism \cite{vaswani2023attentionneed}.  In the MIMO context, each transmit layer can be viewed as a “token” whose mutual interference patterns resemble the contextual relationships among words in a sentence.  A carefully tailored attention mechanism could thus enable a receiver to learn and mitigate inter‐stream interference in a flexible, data‐driven manner, while still preserving interpretability through physically motivated embeddings.

In this paper, we introduce \emph{AttDet}, an AI‐based MIMO detection framework inspired by the Transformer model.  We cast the MIMO detection problem as a sequence prediction task, where the attention scores between stream tokens are directly computed from learned projections of the estimated channel matrices, thereby encoding channel orthogonality into the network’s similarity metric.  By initializing the value vectors with a matched‐filter output and iteratively refining them through several attention layers, AttDet generalizes classical gradient‐based and MMSE updates into a learnable architecture.  We demonstrate through extensive link‐level simulations under 3GPP defined 5G channel models \cite{3gpp_tr_38_901}, high‐order quadrature amplitude modulation (QAM), and both SU-MIMO (Single User) and MU-MIMO (Multi-User) scenarios—that AttDet achieves near‐optimal BER/BLER performance with tractable complexity.  

Our key contributions are:
\begin{itemize}
  \item \textbf{Sequence‐based MIMO detection:} Reformulating the MIMO equalization task as a Transformer‐style self‐attention problem, treating transmit layers as tokens and interference as attention weights.
  \item \textbf{Physically motivated embeddings:} Deriving queries and keys directly from the estimated channel matrices, so that attention scores quantify cross-layer channel correlation and interference.
  \item \textbf{Near‐optimal performance:} Demonstrating that AttDet matches or closely approaches K-best and ML‐based detectors in realistic link‐level simulations, while offering a predictable computational profile suitable for implementation.
\end{itemize}

The remainder of the paper is organized as follows.  Section~\ref{sec:discussion} describes the system model, the relevant prior-art AI-based detectors and classical baselines, Section~\ref{sec:attdet} details the AttDet architecture, Section~\ref{sec:results} presents the link‐level evaluation results, and Section~\ref{sec:conclusions} concludes the paper.

\section{Discussion}
\label{sec:discussion}

\subsection{System Model}
We assume an OFDM based radio link with a configuration inherited from typical 5G system parameters, where there are $N_{sc}$ number of sub-carriers and $N_{symb}$ number of OFDM symbols in one OFDM transmission block, also called a sub-frame or a slot in 4G/5G terminology. The smallest unit in the time-frequency grid of one sub-carrier during one OFDM symbol is called the Resource Element (RE), which carries one modulation symbol.

We consider $N_t$ transmit antennas, which may be associated with one or multiple transmitters (i.e., UEs) and $N_r$ receive antennas at the network side, which essentially forms an uplink MIMO system. We note, however, that our proposed solution remains applicable for the downlink as well.

For the received vector $\textbf{y}\in\mathbb{C}^{N_r}$ the following equation holds
\begin{equation}\label{eq:channel}
  \mathbf{y}_{\left[\mathit{sc},\mathit{sy}\right]} = \mathbf{H}_{\left[\mathit{sc},\mathit{sy}\right]} \mathbf{x}_{\left[\mathit{sc},\mathit{sy}\right]} + \mathbf{n}
\end{equation}
where $\mathbf{H_{\left[\mathit{sc},\mathit{sy}\right]}}\in \mathbb{C}^{N_r\times N_t}$ is the channel matrix, $\mathbf{n}\sim\mathcal{C}\mathcal{N}(0,\sigma^2\mathbf{I}_{N_r})$ is complex Gaussian noise, assuming spatially white noise across the receiver antennas and $\mathbf{x_{\left[\mathit{sc},\mathit{sy}\right]}} \in\mathcal{X}^{N_t}$ is the vector of transmitted symbols, where $\mathcal{X}$ denotes the finite set of constellation points. The equation is per RE, hence the subscripts of $\left[\mathit{sc},\mathit{sy}\right]$ for sub-carrier $\mathit{sc}$ and OFDM symbol $\mathit{sy}$. For simplicity, we omit the use of the subscripts in the rest of the equations.

%
Note that in the above equation we assumed that $N_{t}$ number of MIMO layers ($N_{l}$) are transmitted (i.e., $N_{l}=N_{t}$) but in the general case of ($N_{l} \leq N_{t}$) a pre-coder matrix $\mathbf{P} \in \mathbb{C}^{N_t\times N_l}$ may be applied on $\mathbf{x} \in \mathcal{X}^{N_l}$. The effective channel in (\ref{eq:channel}) becomes $\mathbf{H} \cdot \mathbf{P}$ and the same model remains applicable. 

Note also that the true value of $\textbf{H}$ is not known in any practical system, only an estimate $\hat{\textbf{H}}$, obtained in the channel estimation phase. We assume that for each MIMO layer we choose a symbol randomly from $\mathcal{X}$ according to uniform distribution, and all layers use the same constellation set. Further, as is standard practice, we assume that the constellation set $\mathcal{X}$ is given by a QAM scheme and all constellations are normalized to unit average power.

The goal of the receiver is to obtain the estimate $\hat{\textbf{x}}$ of $\textbf{x}$. The optimal receiver under the assumption of Gaussian noise is the Maximum Likelihood receiver, which solves the following optimization problem:
\begin{equation}\label{eq:problem}
    \hat{\textbf{x}}= \argmin_{\textbf{x}\in \mathcal{X}^{N_t}} ||\textbf{y}-\textbf{Hx}||_2.
\end{equation}

Since the ML receiver scales as $O(M^{N_l})$, where $M$ is the cardinality of the set of allowed modulation symbols, it quickly becomes computationally infeasible.

Linear receivers are practically more feasible as they obtain an estimate of the transmitted symbols as a linear combination of the elements of the received signal vector $\textbf{y}$, i.e., $\hat{\mathbf{x}} = \mathbf{W}\mathbf{y}$. Two of the well-known linear equalizers are ZF and MMSE.

The MMSE equalizer minimizes the mean squared estimation error of
$\textbf{e} = \argmin_{\mathbf{W}\in \mathbb{C}^{N_l \times N_r}} \mathbb{E}\|\mathbf{x} - \mathbf{W} \mathbf{y}\|_2^2$

The solution of the MMSE filter is:
\begin{equation}
\mathbf{W}_{\text{MMSE}} = \left(\mathbf{H}^{H} \mathbf{H} + \sigma^2 \mathbf{I}\right)^{-1} \mathbf{H}^{H}
\end{equation}
where $\mathbf{H}^H$ is the Hermitian transpose of $\mathbf{H}$.

The ZF equalizer simply employ the pseudo inverse of $H$ as the equalization filter, i.e., 
\begin{equation}
\mathbf{W}_{\text{ZF}} = \left(\mathbf{H}^{H} \mathbf{H}\right)^{-1} \mathbf{H}^{H}.
\end{equation}

As these linear methods involve matrix inversion, their compute complexity increases drastically with the number of receive antennas and they fall behind the optimal solution in case of correlated channels.

\subsection{Prior AI-Based MIMO Detection Approaches}

\paragraph{Hybrid Model-Based Detectors} Early works applied deep learning by \emph{unfolding classical algorithms} for MIMO detection. For example, DetNet \cite{samuel2017DeepMIMO,song2019LearnToDetect} and related models unrolled iterative optimization (gradient descent or message passing) into a neural network. Samuel \emph{et al.} (2017) \cite{samuel2017DeepMIMO} introduced DeepMIMO/DetNet by iteratively updating symbol estimates with learnable parameters, effectively learning a projected gradient descent routine:
\begin{equation}
\hat{\mathbf{x}}_{k+1} = \Pi\bigl[\hat{\mathbf{x}}_k - \delta_k \mathbf{H}^H\mathbf{y}
+ \delta_k \mathbf{H}^H \mathbf{H} \hat{\mathbf{x}}_k \bigr] \,,
\label{eq:grad_update}
\end{equation}
where $\Pi[\cdot]$ is a nonlinear projection operator and $\delta_k$ is the step size. Subsequent works like MMNet \cite{shirkoohi2019adaptMassiveMIMO} and DeEQ \cite{hummert2020DeEQ} refined this idea with learned denoisers and parameter sharing, demonstrating gains not just on i.i.d. Gaussian channels but also with more realistic channel models.

\paragraph{Black-Box Neural Detectors} More recent studies treat the MIMO receiver as a \emph{pure data-driven} problem, using deep networks to map directly from received signals to detected outputs. Cammerer \emph{et al.} (2023) \cite{cammerer2023neuralreceiver5gnr} and Korpi \emph{et al.} (2020) \cite{korpi2020deeprxmimoconvolutionalmimo} proposed “DeepRx” architectures that replace entire receiver blocks (channel estimation, equalization, demapping) with CNNs or fully-connected nets. These end-to-end models often outperform MMSE detectors, but have large computation complexity and blend the contribution of channel estimation and MIMO detection in the overall gain.

\paragraph{Graph Neural Network Detectors} To better exploit the MIMO structure, some works employ \emph{graph neural networks} to mimic belief propagation on the detection factor graph \cite{scotti2020graphneuralnetworksmassive}. By representing each transmit layer as a graph node and learned messages as interference cancellation steps, GNN-based detectors can outperform MMSE and standard belief propagation. Cammerer \emph{et al.} (2023) \cite{cammerer2023neuralreceiver5gnr} combined a CNN with a GNN in a universal MU-MIMO receiver, achieving performance close to the K-best detector.

\subsection{Attention Mechanisms in MIMO Detection}

Attention-based architectures have only \emph{recently} been applied to wireless receivers. Burera \emph{et al.} (2025) employ a Transformer encoder block for MIMO detection, capturing global dependencies in the antenna array and achieving significant BER improvements over conventional methods \cite{burera2025transformerMIMO}. Michon \emph{et al.} (2022) \cite{michon2022convSA} proposed a convolutional self-attention demapper for MU-MIMO OFDM, where a CNN learns time–frequency error correlations and a self-attention mechanism weighs residual inter-user interference.

\section{AttDet: Transformer-Inspired MIMO Detection}
\label{sec:attdet}

\begin{figure}[ht]
\centering
\includegraphics[width=0.85\columnwidth]{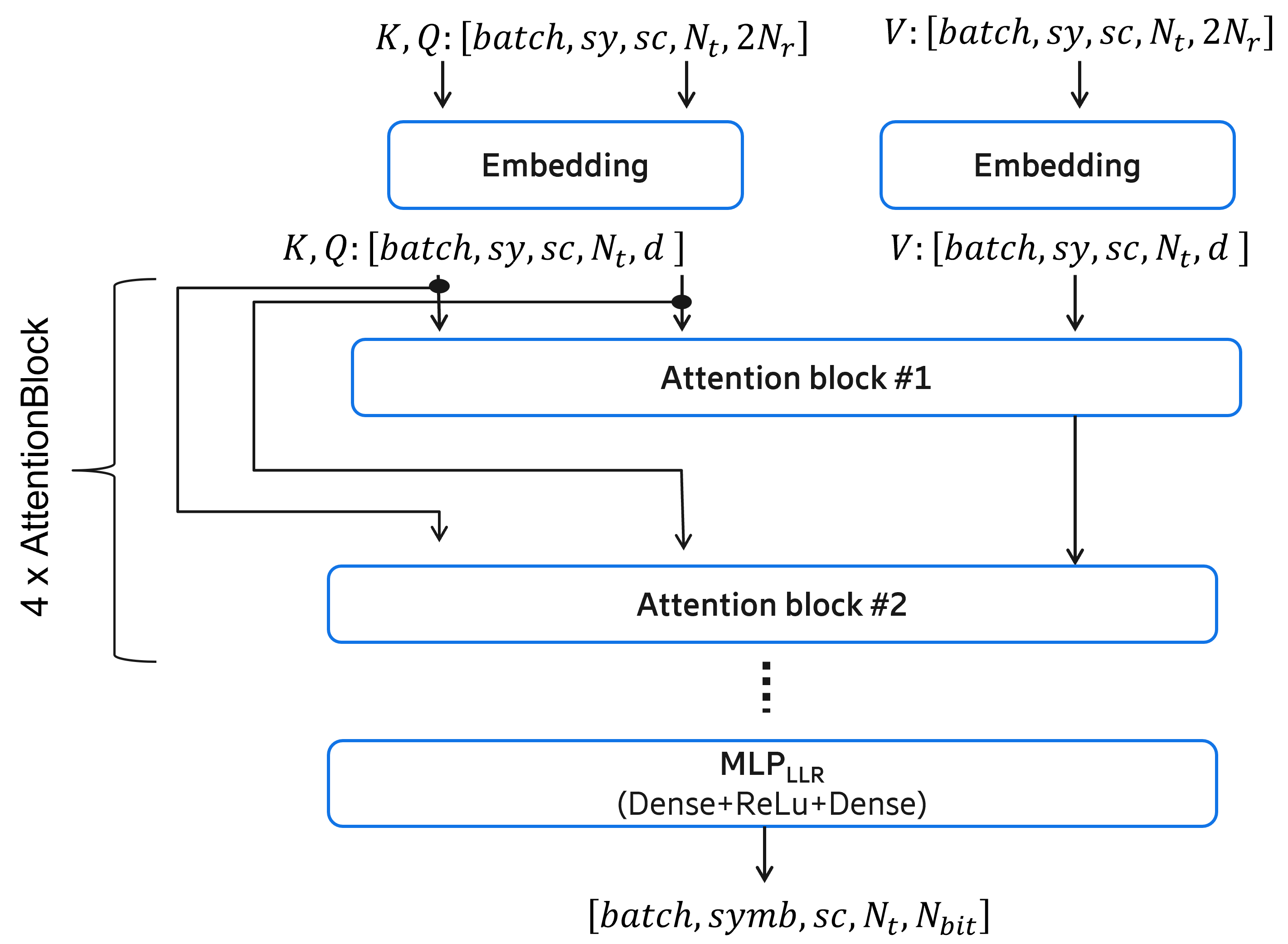}
\caption{AttDet: overall neural architecture}
\label{fig:attdet_arch}
\end{figure}

In this section, we describe our proposed AttDet architecture, which casts MIMO equalization as a self-attention problem. By treating each transmit layer as a “token” and encoding inter-stream interference via learned attention weights, AttDet unifies model-based insight with the flexibility of data-driven optimization. The AttDet neural architecture is illustrated in Figure~\ref{fig:attdet_arch}. 

\subsection{Token Embeddings}

Given an estimated channel matrix $\widehat{\mathbf{H}}\in\mathbb{C}^{N_r\times N_t}$ and receive vector $\mathbf{y}\in\mathbb{C}^{N_r}$, we associate each transmit layer $i=1,\dots,N_t$ with three embeddings:

\begin{align}
\mathbf{q}_i &= \mathrm{MLP}_{QK}\left(\phi\left(\widehat{\mathbf{H}}_{:,i} \right)\right) & \in \mathbb{R}^{d}\label{eq:query}\\
\mathbf{k}_i &= \mathrm{MLP}_{QK}\left(\phi\left(\widehat{\mathbf{H}}_{:,i} \right)\right) & \in \mathbb{R}^{d} \label{eq:key}\\
\mathbf{v}_i &= \mathrm{MLP}_V\left(\phi\left(\frac{1}{\lVert\widehat{\mathbf{H}}_{:,i}\rVert_2^2} \odot \widehat{\mathbf{H}}_{:,i}^* \odot \mathbf{y} \right)\right) & \in \mathbb{R}^{d} \label{eq:value}
\end{align}

where $\phi(\widehat{\mathbf{H}}_{:,i}) \in \mathbb{R}^{2N_r} \text{ be } [\mathrm{Re}(\widehat{\mathbf{H}}_{:,i}),\mathrm{Im}(\widehat{\mathbf{H}}_{:,i})]$ is the operator that flattens the complex valued channel vector as the concatenation of real and imaginary components. The operator $\odot$ means element-wise vector multiplication and $\mathrm{MLP}_{QK}$, $\mathrm{MLP}_V$ are two-layer feed-forward networks and $d$ is the model inner dimension.

Intuitively, the query ($\mathbf{q}_i$) and key ($\mathbf{k}_i$) representations of layer $i$ are derived from the channel vector of layer $i$, while the value $v_i$ is the representation of the transmitted symbol by layer $i$. During the attention layers the value of layer $i$ gets updated with the value of other layers depending on their channel "similarity", acting like an iterative interference compensation layer-by-layer.

\subsection{Attention Layer}

\begin{figure}[ht]
\centering
\includegraphics[width=0.6\columnwidth]{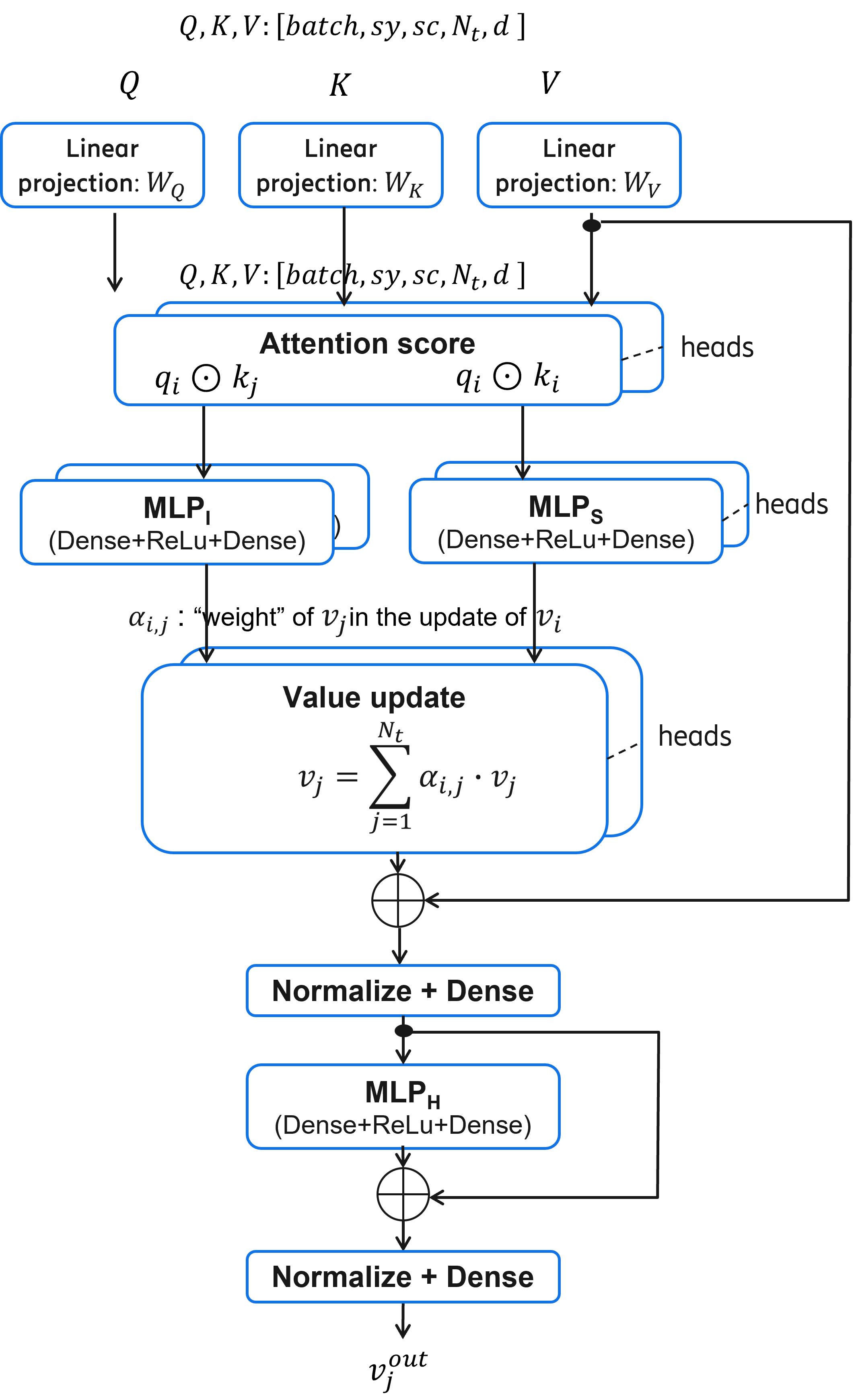}
\caption{Breakdown of the Attention block}
\label{fig:att_block}
\end{figure}

Each attention layer is composed of a (I) linear projection, (II) a similarity calculation and (III) an update of the per-token value vectors according to the similarity metric (see Figure~\ref{fig:att_block}).
At each of $T$ attention layers, we compute pairwise ``interference weights'' between the MIMO layers and refine the value vectors. The scaled dot-product attention score from token $i$ to token $j$ is
\begin{equation}
\alpha_{ij} = \mathrm{MLP}_I\bigl(\mathbf{q}_i \odot \mathbf{k}_j\bigr) \quad i \neq j, \, \text{and}
\label{eq:alpha_ij}
\end{equation}
\begin{equation}
\alpha_{ii} = \mathrm{MLP}_S\bigl(\mathbf{q}_i \odot \mathbf{k}_i\bigr) \quad  i=j,
\label{eq:alpha_ii}
\end{equation}
where $\mathrm{MLP}_I(\cdot)$ and $\mathrm{MLP}_S(\cdot)$ are two-layer feed-forward networks with a ReLu layer in-between and with inner dimension of $4 \cdot d/N_{head}$, where $N_{head}$ is the number of heads in the Attention blocks (see later). The $\mathrm{MLP}_I$ network learns the vector transformation to be applied on the cross-layer interference, while $\mathrm{MLP}_S$ learns the transformation to be applied on the wanted signal during the value update.

These weights encode channel orthogonality, i.e., strongly correlated layers receive higher mutual attention. We then update each value vector by 
\begin{equation}
\mathbf{v}_i^{(t+1)}
=\sum_{j=1}\alpha_{ij}\,\odot \mathbf{v}_j^{(t)},
\quad t=0,\dots,T-1.
\label{eq:value_update}
\end{equation}
This update parallels classical successive-interference-cancellation and MMSE refinements, with the MLPs providing nonlinear symbol correction.

The final step in the Attention block is a two-layer feed-forward network with ReLu layer in-between with inner dimension of $4 \cdot d$, which is applied on the updated value vectors individually:
\begin{equation}
\mathbf{v}_i^{(out)} = \mathrm{MLP}_{H}\bigl(\mathbf{v}_i^{(t+1)}\bigl).
\label{eq:mlp_heads}
\end{equation}
The purpose of the $MLP_H$ layer is also to mix the outputs of multiple heads, in case there are more than one head in the Attention layer. In case of multiple heads in the Attention block the attention score and value update layers are performed via separate MLPs per-head and the query, key, value vector dimensions are split across the heads, so that each head deals with $d/N_{head}$ dimensional token vectors. According to our observation, applying multiple heads does not improve performance but reduces model size and computational complexity. Setting $N_{head}=4\dots8$ keeps a reasonably sized per-head dimension, maintains performance and reduces computation at the same time.

We have observed that introducing a convolution layer on the similarity scores before applying the MLP in (\ref{eq:alpha_ij}) and (\ref{eq:alpha_ii}) was able to improve the performance. More specifically, we employ a depth-wise separable convolution of size $3 \times 3$ over the time-frequency dimensions. The convolution enables to add a ``smoothing'' effect of the similarity scores across the neighbouring REs.

\subsection{Output and Decision}

After $T$ layers, the final value embeddings $\{\mathbf{v}_i^{(T)}\}$ are projected back to bit logits:
\begin{equation}
\hat{\boldsymbol{\ell}}_i
= \mathrm{MLP}_{LLR}\bigl(\mathbf{v}_i^{(T)}\bigr) \in\mathbb{R}^{\log_2|\mathcal{X}|},
\end{equation}
where $\mathrm{MLP}_{LLR}(\cdot)$ is a two layer, feed-forward network with a ReLu in-between, computing the Log-Likelihood Ratio (LLR) per bit. We note that the model can be trained for multiple modulation orders either by branching in the last layer of $\mathrm{MLP}_{LLR}$ according to QAM order or by masking the relevant bits on the output of $\mathrm{MLP}_{LLR}$ when calculating the loss function.

\subsection{Remarks}
It is interesting to observe some intuitive analogy between the update equation of AttDet in (\ref{eq:value_update}) and the iterative gradient descent based optimization in (\ref{eq:grad_update}). Note that the attention scores of $\alpha_{ij}$ derived from the query and keys products, which are originating from $\mathbf{H}$ via learnt embeddings are analogous to the $\mathbf{H}^T \mathbf{H}$ term in (\ref{eq:grad_update}). The second term in (\ref{eq:grad_update}) is captured in the initialization of the value vectors in (\ref{eq:value}), while the first term ($\mathbf{x}_k$) corresponds to the skip connection applied around the attention block (see Figure~\ref{fig:att_block}). 
The projection operation in (\ref{eq:grad_update}) is replaced by $\mathrm{MLP}_{H}$ in (\ref{eq:mlp_heads}).

Some key properties and benefits of the AttDet architecture include
\begin{itemize}
  \item \textbf{Physical interpretability:} Queries and keys are directly derived from the estimated channel matrix columns, so attention scores have a clear meaning in terms of inter-stream correlation.
  \item \textbf{Bridging physical models with sound AI architecture:} Giving more freedom to learn within physical inspired constraints built into the custom Transformer architecture.
  \item \textbf{Isolation of detection:} By feeding in only the channel estimate and receive vector, we focus training on equalization, avoiding coupling with end-to-end channel-estimation errors. 
  \item \textbf{Generalization:} The same architecture adapts seamlessly to SU- and MU-MIMO settings, different $N_t,N_r$, and arbitrary QAM orders, as we are going to show in the evaluations.
\end{itemize}

\section{Evaluations}
\label{sec:results}

\subsection{Simulation parameters}
We use link level simulations for the evaluation with 3GPP UMa channel model, the Modulation and Coding Scheme (MCS) has been fixed at some selected operating points according to the modulation order (MCS=5 for QPSK, MCS=13 for 16QAM and MCS=20 for 64QAM). The operating frequency is set to 3.7 GHz, the OFDM sub-carrier spacing is 30 kHz and the whole bandwidth includes 25 PRBs (Physical Resource Block). The UE speed varied between 3-37 m/sec during training and was fixed to 3 m/sec at evaluation. The number of antenna elements at the receiver was $(1 \times 4 \times 2)$ in $(\text{vert} \times \text{horiz} \times \text{pol})$ dimensions, unless specified otherwise. The UE has one single antenna element in the MU-MIMO case and two, co-polarized elements in the SU-MIMO case. We note that in many studies it is generally assumed that UE antennas are cross-polarized, however, in practice it is typically a mixture of the co- and cross-polarized effects due to the many imperfections. Therefore, in our evaluations we assume the worst case, co-polarized correlation between the MIMO channels. We run SU-MIMO evaluations with 2 layers and 1 UE and MU-MIMO with the number of users varied between 2-4 and each transmitting one layer. We use these settings as typical parameters of realistic low-band deployments.

Since the channel estimation is not part of the AI model, we employ an MMSE based method to obtain initial channel estimates that are input to the model.

We run the training for approx. 6 million distinct samples and train the model in a supervised way against the binary cross-entropy with the ground truth transmitted bit. We do the evaluation on an independently generated data set.

\subsection{SU-MIMO}
\begin{figure}
\centering
\includegraphics[width=0.35\textwidth]{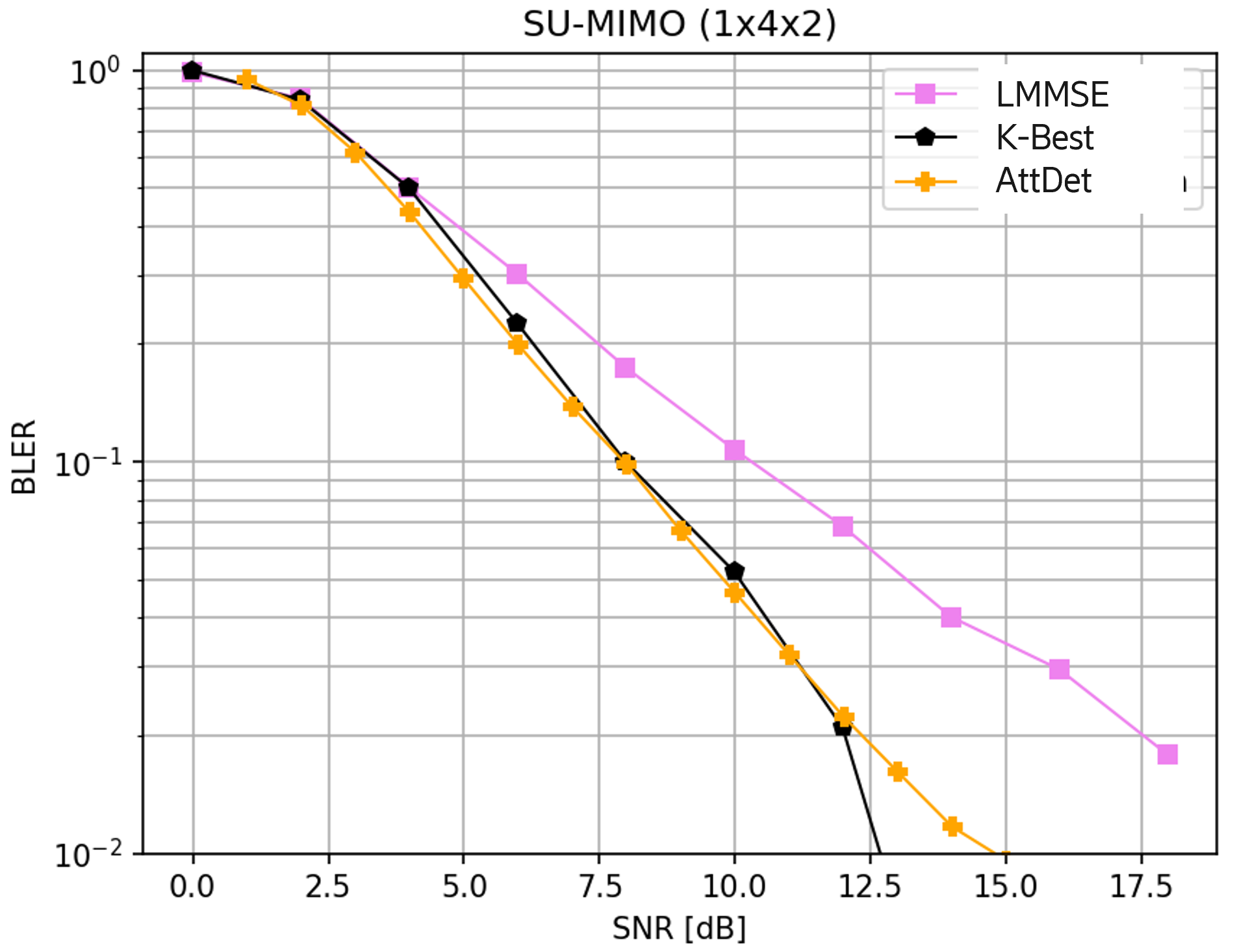}
\caption{\label{fig:bler_sumimo_qam16}BLER for SU-MIMO, $N_r=8$, $N_l=2$, 16QAM}
\end{figure}

The BLER performance of the SU-MIMO case for 16QAM is shown in Figure~\ref{fig:bler_sumimo_qam16}. As the channels of the MIMO layers are correlated stronger in SU-MIMO (due to the tx antennas being at the same location and being co-polarized) than in MU-MIMO, the detection is hard to solve optimally with a linear method. Hence there is a relatively large gap and potential gain between LMMSE and K-Best (K=64) but AttDet can fully realize this gain and achieve the quasi ideal performance. At the operation point of interest of $\sim\negmedspace 0.1$ BLER, this gain is around $\sim\!2\text{--}3$ dB.

\begin{figure}
\centering
\includegraphics[width=0.35\textwidth]{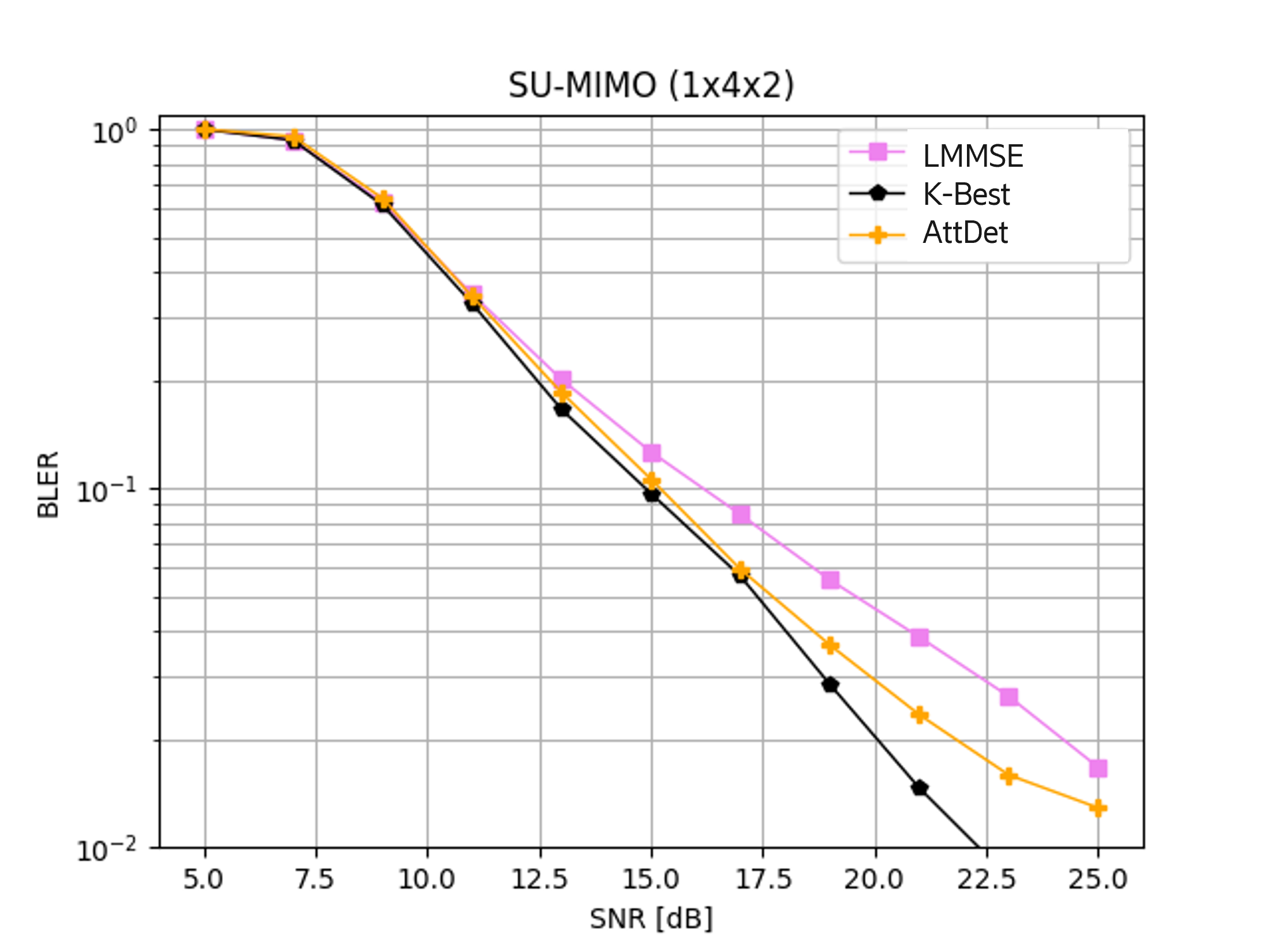}
\caption{\label{fig:bler_sumimo_qam64}BLER for SU-MIMO, $N_r=8$, $N_l=2$, 64QAM
}
\end{figure}

As the detection problem scales, e.g., by increasing the modulation order or the number of layers, the gap and the potential gain between LMMSE and K-Best (K=256) is decreasing but AttDet can still realize this maximum possible gain, see Figure~\ref{fig:bler_sumimo_qam64} SU-MIMO results for the 64QAM case, where there is still approx. $\sim\!1\text{--}1.5$ dB gain.

\begin{figure}
\centering
\includegraphics[width=0.35\textwidth]{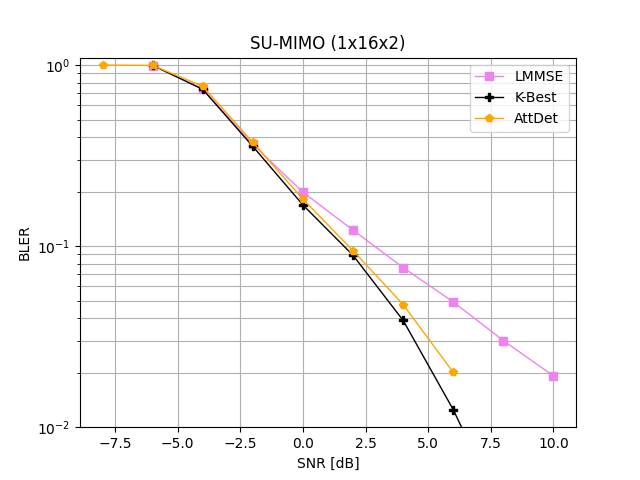}
\caption{\label{fig:bler_sumimo_rx32}BLER for SU-MIMO, $N_r=32$, $N_l=2$, 16QAM
}
\end{figure}

It is also interesting to observe that by increasing the number of receive antenna elements to 32 (see Figure~\ref{fig:bler_sumimo_rx32}) and thereby increasing the potential ``separability'' of the layers does not help the LMMSE detector to better approach K-Best performance, while AttDet can still perform close to K-Best. The reason is probably that the channels are correlated already at the tx side and it remains hard to separate them with linear methods at the receiver.

\begin{figure}
\centering
\includegraphics[width=0.35\textwidth]{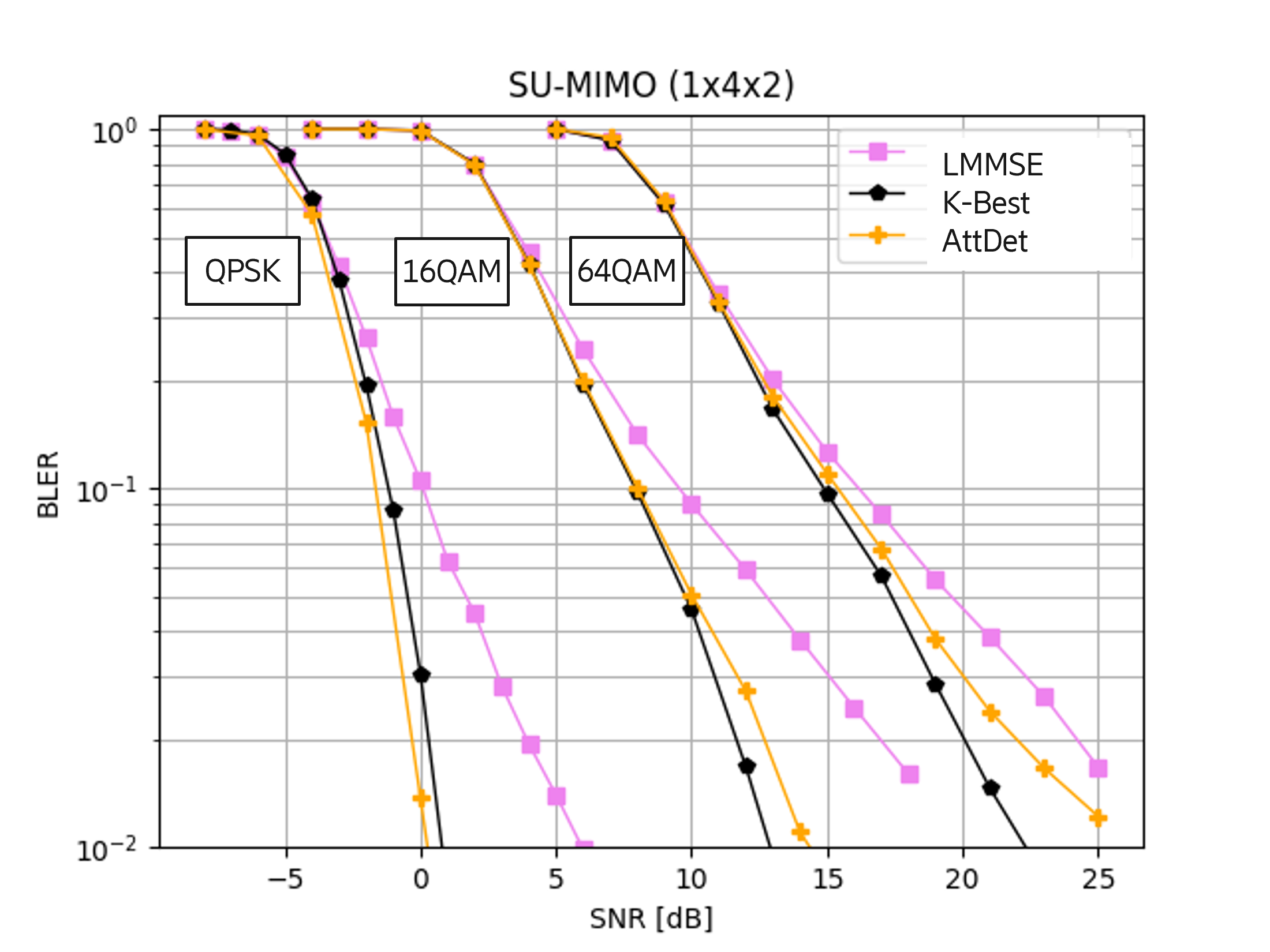}
\caption{\label{fig:bler_sumimo_multi_mcs}BLER for SU-MIMO, $N_r=8$, $N_l=2$, multi-QAM
}
\end{figure}

Next, we trained a model with samples of different modulations (but no mixing of modulation orders for the layers within the same sample), including QPSK, 16QAM and 64QAM constellations and tested for each case separately, as shown in Figure~\ref{fig:bler_sumimo_multi_mcs}. The performance gains remain the same with the mixed model, meaning that a single model is sufficient to cover multi-MCS configurations. In this case the model is trained for a mix of different modulation orders. We note that the same model generalizes also for multiple layers, as new layers appear as additional tokens in the sequence but the model and its parameters can remain the same.

\subsection{MU-MIMO}
\begin{figure}
\centering
\includegraphics[width=0.35\textwidth]{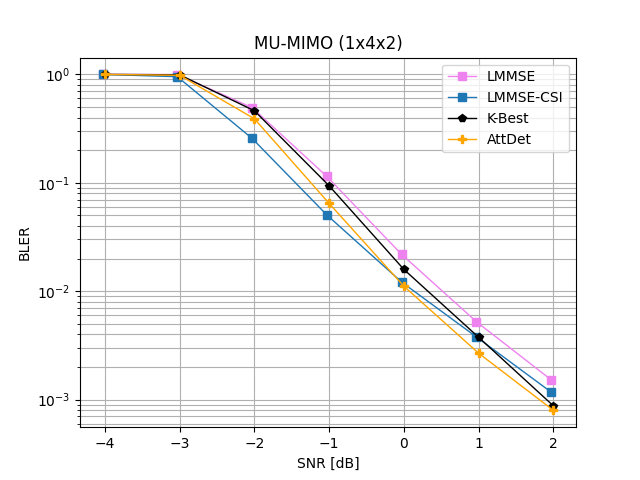}
\caption{\label{fig:bler_mumimo_ue2_qam16}BLER for MU-MIMO, $N_r=8$, $N_l=2$, 16QAM
}
\end{figure}

Due to the  diversity of UE locations, the MIMO channels are less correlated in the MU-MIMO case, hence the potential gains are also smaller. In Figure~\ref{fig:bler_mumimo_ue2_qam16} we plot the 2 UE case with 16QAM and 8 rx antennas.
We also plot the LMMSE detector with ideal channel state information (CSI) as a reference to see the potential gains when channel estimation is not limiting. We can observe that the AttDet receiver (with practical channel estimation) can still approach LMMSE performance with ideal channel knowledge. 

\begin{figure}
\centering
\includegraphics[width=0.35\textwidth]{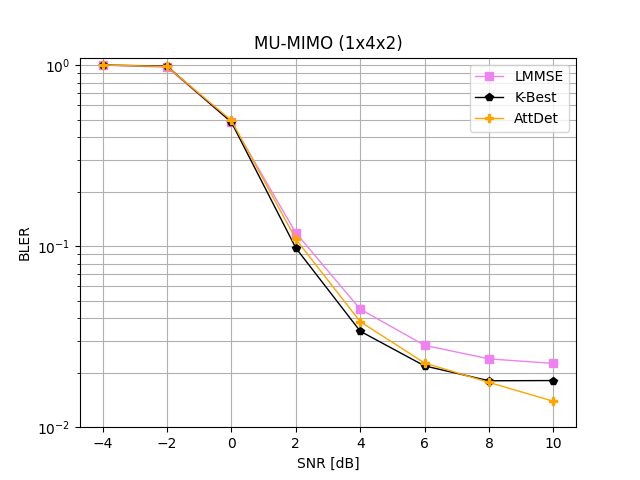}
\caption{\label{fig:bler_mumimo_ue4_qam16}BLER for MU-MIMO, $N_r=8$, $N_l=4$, 16QAM
}
\end{figure}

Then we increase the number of UEs to 4 in Figure~\ref{fig:bler_mumimo_ue4_qam16} and observe a diminishing potential of gains, which is most probably due to getting closer to the physical limitations of layer separability as $N_l \to N_r$.
all three detectors performing roughly the same (at least in the operating region of interest) but AttDet still outperforms LMMSE.

\section{Conclusions}
\label{sec:conclusions}
In this work, we have introduced AttDet, a novel MIMO detection architecture that reformulates the equalization task as a Transformer-style self-attention problem. By utilizing physically motivated embeddings that derive queries and keys directly from estimated channel matrices and value update operations that mimic MIMO equalization steps, our approach bridges classical MIMO detection techniques with modern AI architectures.

Extensive link-level simulations demonstrate that AttDet achieves near-optimal detection performance, closely matching advanced baselines such as K-best, under practical channel models and channel estimation conditions. In scenarios with correlated MIMO channels (e.g., in SU-MIMO) there is a significant gain potential and gap between classical linear detector (LMMSE) and K-Best and AttDet is able to realize these gains.

Its complexity remains polynomial and scales quadratically in the number of transmitted layers but remains independent of the number of receive antennas or modulation order. This is an obvious advantage over high-complexity tree-search of K-Best, scaling exponentially with modulation order and also over LMMSE requiring an antenna size dependent matrix inversion. We note, however, that more work would be needed to investigate complexity reduction options either via architecture re-design or pruning techniques.

\section{Acknowledgements}
The authors would like to thank Karl Werner, M{\aa}rten Sundberg and Shashi Kant, all are at Ericsson for their insightful discussions and valuable revision suggestions, which significantly enhanced this work. 

\bibliographystyle{alpha}
\bibliography{MIMO_AI_receiver_arxiv_submit.bib}

\end{document}